\documentclass[%
 aip,
 amsmath,amssymb,
 reprint,%
]{revtex4-2}

\usepackage{graphicx}
\usepackage{dcolumn}
\usepackage{bm}
\usepackage{natmove}

\usepackage[utf8]{inputenc}
\usepackage[T1]{fontenc}
\usepackage{mathptmx}
\usepackage{soul}
\usepackage{tikz}
\usepackage{xcolor}

\usepackage[normalem]{ulem} 

\newcommand\rout{\bgroup\markoverwith
{\textcolor{red}{\rule[.5ex]{2pt}{2pt}}}\ULon}

\newcommand{\kT}{k_{\rm B} T}

\newcommand{\bG}{{\bf G}}

\newcommand{\bphi}{{\boldsymbol \phi}}
\newcommand{\bb}{{\boldsymbol b}}

\begin{document}

\title{Broad chemical transferability in structure-based coarse-graining }

\author{Kiran H.~Kanekal}
\email{kkanekal@gmail.com}
\affiliation{Max Planck Institute for Polymer Research, Ackermannweg 10, 55128 Mainz, Germany
}%

\author{Joseph F.~Rudzinski}
\affiliation{Max Planck Institute for Polymer Research, Ackermannweg 10, 55128 Mainz, Germany
}%

\author{Tristan Bereau}
\affiliation{Van ’t Hoff Institute for Molecular Sciences and Informatics Institute, University of Amsterdam, Amsterdam 1098 XH, The Netherlands
}%
\affiliation{Max Planck Institute for Polymer Research, Ackermannweg 10, 55128 Mainz, Germany
}%

\date{\today}

\begin{abstract}
  Compared to top-down coarse-grained (CG) models, bottom-up
  approaches are capable of offering higher structural fidelity. This
  fidelity results from the tight link to a higher-resolution
  reference, making the CG model  chemically specific. Unfortunately,
  chemical specificity can be at odds with compound-screening
  strategies, which call for transferable parametrizations. Here we
  present an approach to reconcile bottom-up, structure-preserving CG
  models with chemical transferability. We consider the bottom-up CG
  parametrization of 3,441 C$_7$O$_2$ small-molecule isomers. Our
  approach combines atomic representations, unsupervised learning, and
  a large-scale extended-ensemble force-matching parametrization. We
  first identify a subset of 19 representative molecules, which
  maximally encode the local environment of all gas-phase conformers.
  Reference interactions between the 19 representative molecules were
  obtained from both homogeneous bulk liquids and various binary
  mixtures. An extended-ensemble parametrization over all 703 state
  points  leads to a CG model that is both structure-based and
  chemically transferable. Remarkably, the resulting force field is on
  average more structurally accurate than single-state-point
  equivalents. Averaging over the extended ensemble acts as a
  mean-force regularizer, smoothing out both force and structural
  correlations that are overly specific to a single state point. Our
  approach aims at transferability through a set of CG bead types that
  can be used to easily construct new molecules, while retaining the
  benefits of a structure-based parametrization.
\end{abstract}

\maketitle

\section{Introduction}

In order to facilitate molecular design for a wide variety of
applications, there has recently been a growing interest in utilizing
data-driven techniques to infer chemical structure--property
relationships that span broad regions of chemical compound space (CCS)
\cite{kuhn1996inverse, bereau2016research, ramprasad2017machine,
sanchez2018inverse, sherman2020inverse}. A common rate-limiting step
in deriving these relationships is acquiring target properties for a
sufficient number of compounds, so as to ensure robustness and
transferability. As such, a push for increasingly automated workflows
for generating data via both experimental and computational methods
has risen in tandem with these data-driven approaches. While
experimental approaches are limited due to material cost and ease of
chemical synthesis, computational methods do not suffer from these
restrictions. Instead, computation is primarily limited by sampling,
calling for ever-improving high-performance computing platforms or
algorithms \cite{shirts2000screen, giupponi2008impact, shaw2008anton}.
The limitations to computational high-throughput screening often stem
from the prohibitive  computational cost of simulating large systems
(on the order of thousands of atoms) at atomic or electronic
resolutions.\cite{bereau2020computational}

A different strategy to computationally screen across more compounds
consists of relying on lower-resolution models. Here we focus on
particle-based coarse-grained (CG) simulations, in which groups of
atoms are mapped to superparticles or beads
\cite{noid2013perspective}. The interactions that govern the behavior
of these beads aim at recovering the essential physics of the system.
This results in simulations that are more computationally efficient
due to the reduction in number of particles and a smoothened
free-energy landscape. In the context of screening, some CG models
offer even more computational efficiency: the CG representation
averages over \emph{molecules}, easing the coverage of CCS. These CG
models, commonly called transferable, reduce the size of CCS by making
use of a discrete set of CG bead types \cite{bereau2015automated,
kanekal2019resolution}. Transferable CG models have been used to
efficiently cover large subsets of CCS and rapidly sketch
structure--property relationships for complex thermodynamic properties
\cite{menichetti2018drug, hoffmann2019controlled}. These studies
relied on the biomolecular Martini force field, a top-down CG model
aiming to reproduce thermodynamic-partitioning behavior in different
environments \cite{periole2013martini}. While top-down CG models can
prove extremely efficient to parametrize and extend, they often
feature limited structural accuracy \cite{alessandri2019pitfalls}.

To construct structurally accurate CG models, bottom-up methods offer
a more systematic route \cite{tschop1998simulation,
izvekov2005multiscale, shell2008relative}. They derive CG interactions
by matching microscopic information from a higher-resolution
reference, for instance the radial distribution function (RDF) or
other features of the many-body potential of mean force (MBPMF).
The reduction in the number of degrees of freedom makes these target
properties inherently dependent not only on the chemical composition,
but also the thermodynamic state point. It is thus no surprise that
most bottom-up CG studies have focused on individual reference
systems. There are various strategies to build bottom-up CG models
that are state-point and/or chemically transferable. Intuition can go
a long way: different molecules may inspire a consistent CG mapping
and set of bead types. For instance, Wang and Deserno parametrized a
CG model for phospholipid membranes and showed that the same set of CG
beads could be used to construct reliable models for lipids with
different saturation levels \cite{wang2010systematic}. In general
though, intuition may not be a silver bullet, in particular when
bridging across chemical compositions. Van der Vegt and coworkers have
demonstrated that an approach based on thermodynamic cycles can
provide improved thermodynamic and chemical transferability, with
respect to alternative bottom-up methods, subject to the limitations
of the form of the interaction potentials.\cite{Brini:2012,
Brini:2012b, Deichmann:2019} Several groups have used local
density-dependent potentials to derive CG models that are transferable
across binary mixture concentrations and phases, providing a more
accurate description of liquid-vapor interfaces~\cite{DeLyser:2017,
Sanyal:2018, Jin:2018, DeLyser:2019, Rosenberger:2019, Shahidi:2020}.
Sanyal \emph{et al.}~recently developed an extended-ensemble
relative-entropy method and constructed a CG protein-backbone model
that could accurately reproduce the structures of over 200 different
globular proteins \cite{sanyal2019hybrid}. 

Counter to the expectation that a single model can reproduce the
behavior of many different types of systems, transferability may
require defining environment-dependent interactions. ``Ultra
coarse-grained'' models are built from a series of internal states
\cite{dama2013theory}. They can accurately model challenging
liquid--vapor and liquid--liquid interfaces \cite{jin2018ultra}. CG
``conformational surface hopping'' applies a simple tuning of the
state probabilities to transfer CG models across both state points and
chemistry\cite{Bereau:2018, rudzinski2020coarse}. Other approaches
aiming at transferability tend to combine multiple references. For
instance, the extended ensemble framework augments the force-matching
based multiscale coarse-graining (MSCG) method by averaging over
multiple MBPMFs \cite{mullinax2009extended}. Mullinax and Noid applied
the extended-ensemble approach to build CG potentials of alkanes and
alcohols that aim to be transferable across liquid-state binary
mixtures \cite{mullinax2009extended}.

\begin{figure*}[htbp]
  \begin{center}
    \includegraphics[width=0.9\linewidth]{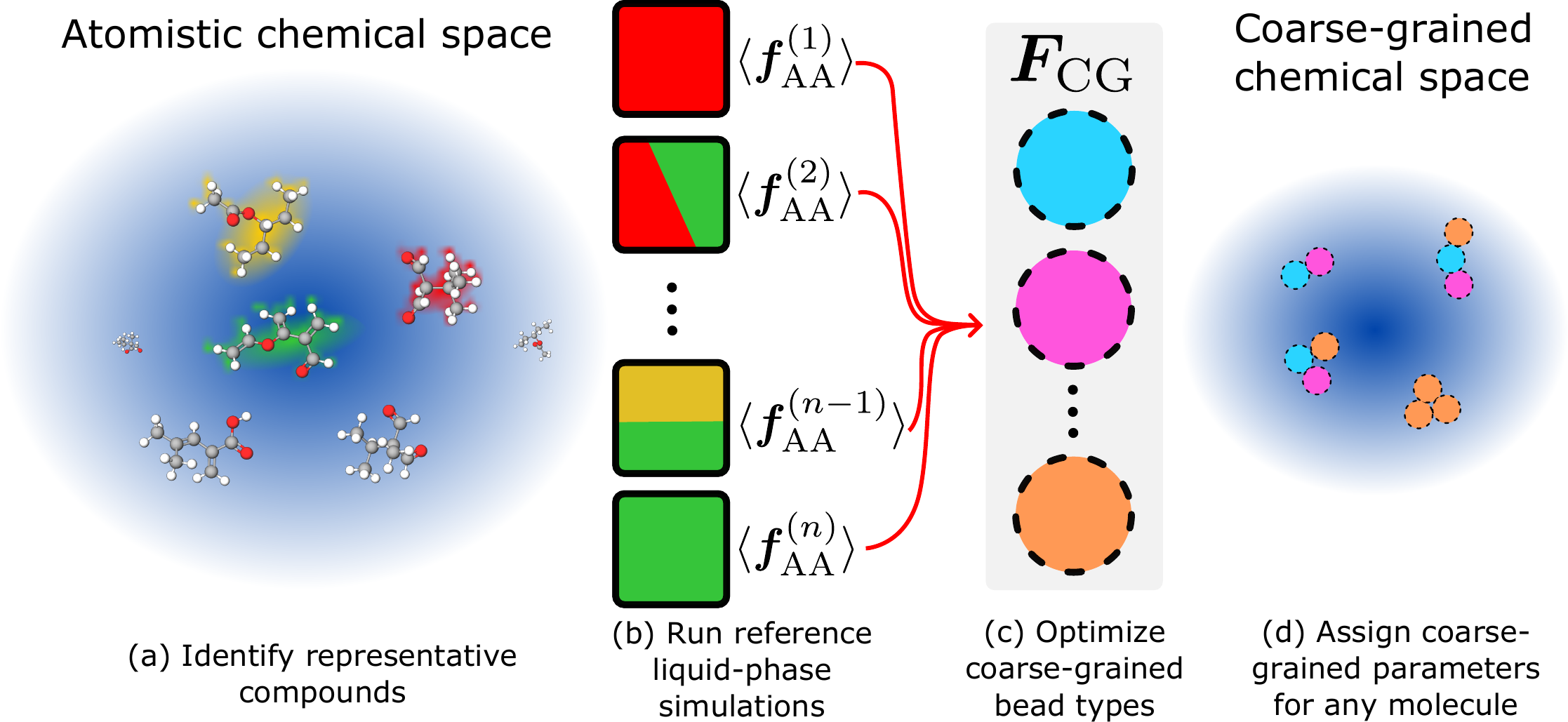}
    \caption{Schematic of our protocol to develop broad chemical
      transferability in a structure-based coarse-grained (CG) model.
      (a) Given an atomistic chemical space, identify representative
      compounds (see Fig.~\ref{fig:ee_umap}); (b) Run reference
      (atomistic) liquid-phase simulations for various homogeneous
      liquids and binary mixtures; (c) Optimize a set of CG bead types
      using an extended-ensemble force-matching scheme; (d) The bead
      types can readily be used to parametrize any molecule in the
      (smaller) CG chemical space.}
    \label{fig:ee_intro}
  \end{center}
\end{figure*}

In this work, we extend the scope of bottom-up CG parametrizations to
target a significantly larger collection of state points and chemical
compositions. Conceptually we seek a CG parametrization scheme that
benefits from multiple reference calculations from various parts of
chemical compound space. We extend the scope of structure-based and
chemically-transferable CG models by simultaneously parametrizing
several thousand small organic molecules---the largest bottom-up CG
parameterization, to the best of our knowledge. Our data-driven and
hierarchical approach is illustrated in Fig.~\ref{fig:ee_intro}. Given
a set of chemical compounds, we first identify a small number of
``representative compounds,'' whose configurational space is best
representative of the entire set. Overall our workflow consists of:
(a) Using gas-phase conformationally averaged many-body atomic
environments, we identify a small number of ``representative''
molecules; (b) Various atomistic simulations of homogeneous liquids
and binary mixtures provide reference mean forces; (c) An
extended-ensemble MSCG method simultaneously parametrizes a force
field with a small collection of CG bead types over all state points;
(d) The set of optimized bead types readily provides nonbonded
parameters for \emph{all} compounds.

We illustrate our approach on 3,441
C\textsubscript{7}O\textsubscript{2} isomers found in the Generated
Database (GDB) \cite{Fink2005,Fink2007}. The identification of 19
representative compounds leads to the generation of 703 atomistic
liquids and binary mixtures, used \emph{simultaneously} to parametrize
our CG model. We then quantify the accuracy of the transferable CG
model by comparing the RDFs to atomistic references. We also benchmark
our transferable model against ``traditional'' state-point-specific CG
force fields.

The results show that enforcing state-point and chemical
transferability in CG potentials can yield high structural accuracy.
Remarkably, the extended-ensemble parametrization is on average
\emph{more accurate} than state-point specific force fields.
Specifically, we find that gains in accuracy are due to a
``regularization-like'' effect that effectively smooths the average
forces acting on specific CG bead types. Averaging over distinct state
points and environments reduces the overfitting of system-specific
features. Similarly, cross correlations inferred from the atomistic
reference simulations are also smoothed, counteracting errors that
arise due to the pairwise form of the CG interactions. On the other
hand, we also find a few examples where the extended-ensemble model
performs notably worse. Low performance stems from certain functional
groups that promote vastly different conformational ensembles
depending on the environment and molecular topology. An
extended-ensemble average over the structural correlations of these
functional groups does not capture the specificity of these diverging
conformational states, and instead suggests the need for an improved
mapping \cite{chakraborty2020preservation, foley2020exploring,
giulini2020information} or an increased force-field complexity
\cite{molinero2009water, john2017many, Bereau:2018,
sanyal2018transferable}. We validate the transferability of the
derived potentials by running CG simulations on compounds that were
not used in the extended-ensemble training set and find that the
accuracy of the CG RDFs is on par with that of the representative
compounds. Overall, we provide a systematic means to perform a
bottom-up coarse-graining over several thousand molecules, resulting
in chemically-transferable CG potentials that retain structural
accuracy in liquid simulations. At the same time, we highlight the
limitations of this approach and note key implementation pitfalls to
avoid.

\section{Methods}

\subsection{Nomenclature}

We first clarify our nomenclature:
\begin{itemize}
  \item We consider the \emph{chemical space} of 3,441
  C\textsubscript{7}O\textsubscript{2} isomers---the entire collection
  of molecules considered.
  \item  Out of the chemical space considered we focus on 23
  molecules. From a clustering analysis, we identify $N_{\rm r}=19$
  \emph{representative compounds}, which are shown in the SI
  (Figs.~S1-S4); 5 additional compounds are selected for
  validation. Each selected compound and CG mapping are denoted by
  numbers, where compound numbers run from 0 to 23 (i.e., 0--18 denote the representative
  compounds, and 19--23 refer to the test compounds). Mapping numbers
  start from 0 and go up to the handful of possibilities e.g.,
  Molecule 21 with Mapping 0. 
  \item Each of the reference compounds is simulated at an atomistic
  resolution in a homogeneous liquid and in all considered binary
  mixtures, leading to $N_{\rm r}(N_{\rm r}+1)/2=190$ reference
  \emph{systems}. Systems only refer to the chemical species; as
  examples, the Molecule 2/Molecule 3 binary system, or the Molecule
  10 pure system can be used to describe any simulation containing
  these particular sets of compounds.
  \item A \emph{state point} denotes the particular thermodynamic
  parameters, including concentration. Specifically, we simulated each
  binary mixture at 4 different concentrations, corresponding to four
  state points per system.
  \item We refer to each combination of system and state point as an
  \emph{ensemble}. The aggregate number of homogeneous liquids and
  binary mixtures of all 19 representative compounds at 4 different
  concentrations amount to a total of 703 atomistic ensembles
  simulated for this work. 
  \item Upon coarse-graining, it does not suffice to define the system
  and state point, but we also need to describe the mapping used, the
  combination of which we refer to as the \emph{mapped ensemble}. A single
  atomistic ensemble may give rise to multiple mapped ensembles, if at
  least one of the compounds has more than one possible CG mapping. In
  this work, the 703 atomistic ensembles translate to a total of 2,476
  mapped ensembles.
\end{itemize}

\subsection{Database}

We selected a subset of the Generated Database (GDB), a computer-generated set of drug-like organic compounds, to test our data-driven bottom-up approach \cite{Fink2005,Fink2007}. Specifically, we selected the set of GDB compounds which were made up of seven carbon atoms and two oxygen atoms only. We further filtered out any compounds containing triple bonds. After applying these filters, we were left with a database of 3,441 C\textsubscript{7}O\textsubscript{2} isomers, listed in their simplified molecular-input line-entry system (SMILES) format. Despite restricting the size of the molecules and only including three elements (C, O, and H), a large variety is still present in the resulting chemical structures. Furthermore, complex interactions, such as hydrogen-bonding and $\pi$-stacking interactions, are also present for many of the compounds in this database. Because the database was limited in terms of the chemical elements, but still contained compounds which we expected to display complex behavior in the bulk phase, we felt this choice of database would prove useful for determining which specific physical interactions would be (un)successfully captured by our chemically-transferable model.

\subsection{Gas-phase simulations}

For each compound in the database, we first ran single-molecule gas-phase
molecular dynamics simulations. The initial structures were obtained by
converting the molecules from their SMILES string representations to
energy-minimized 3D conformations using the \textsc{rdkit} package
\cite{landrum2013rdkit}. The force field parameters for each compound were
generated using the \textsc{cgenff} tool, included in the \textsc{silcsbio} 2018
package, which automatically assigns parameters from the CHARMM General Force
Field based on the input chemistry \cite{vanommeslaeghe2012automation}. The
simulations were run at constant volume using a stochastic velocity-rescaling
thermostat\cite{bussi2007canonical} to maintain a constant temperature,
$T=300~$K. The simulations were run using a 2 fs timestep for a total of 3 ns,
with the LINCS algorithm used to constrain terminal bonds to hydrogen atoms
\cite{hess1997lincs}. A frame was output every 2 ps, yielding 1500 frames per
simulation for each compound in the database. The \textsc{gromacs} 16.1 package
was used to run all of the systems simulated in this work at the atomistic
resolution \cite{abraham2016gromacs}.

\subsection{Defining local environments with SLATM}

The Spectrum of London Axilrod-Teller-Muto (SLATM) vector describes a molecule
as a sum of atomic environments that encode the 1-, 2-, and 3-body interactions
within a cut-off distance \cite{huang2017efficient,Huang2020}. For each atom,
its corresponding SLATM vector consists of the elemental atomic number (1-body),
a spectrum of 2-body London interactions convoluted with a gaussian function
(2-body), and a spectrum of 3-body Axilrod-Teller-Muto interactions also
convoluted with a gaussian function (3-body). The two-body spectrum is computed
over the distance as a London interaction between all pairs within a cut-off
value with a specified step-size. Similarly, the three-body spectrum is computed
as an Axilrod-Teller-Muto interaction over the angle for all triplets within the
cut-off distance.
We applied the \textsc{qml} package made for \textsc{python} 2.7 to convert our
database of compounds into aSLATM representations \cite{christensen2017qml}. The
default values, which were optimized for predicting quantum-mechanical
properties, were used, with a cutoff value of 0.48 nm  and a grid spacing of
0.003 nm and 0.03 radians for the 2-body and 3-body spectra, respectively.

Each frame of the gas-phase simulations yields nine atomic SLATM vectors, i.e.,
one vector per heavy atom, ignoring hydrogens. Because the number of heavy atoms
and chemical composition was constant across the entire database, the length and
ordering of the many-body types for each aSLATM vector was the same.
Fig.~S1 shows the aSLATM vectors of the first
molecule over the entire simulation projected into two dimensions using UMAP
\cite{mcinnes2018umap}. There are only four large clusters due to the symmetry
of the compound. \textsc{hdbscan} facilitated the identification of
clusters in an automated fashion, and seemed relatively insensitive to the choice of \textsc{hdbscan}
parameters \cite{mcinnes2017hdbscan}. The use of different clustering approaches as well as the robustness of the results with respect to the parameters used for these approaches will be the subject of a future study.

\subsection{Selecting representative molecules}

\begin{figure*}[htbp]
  \begin{center}
    \includegraphics[width=0.8\linewidth]{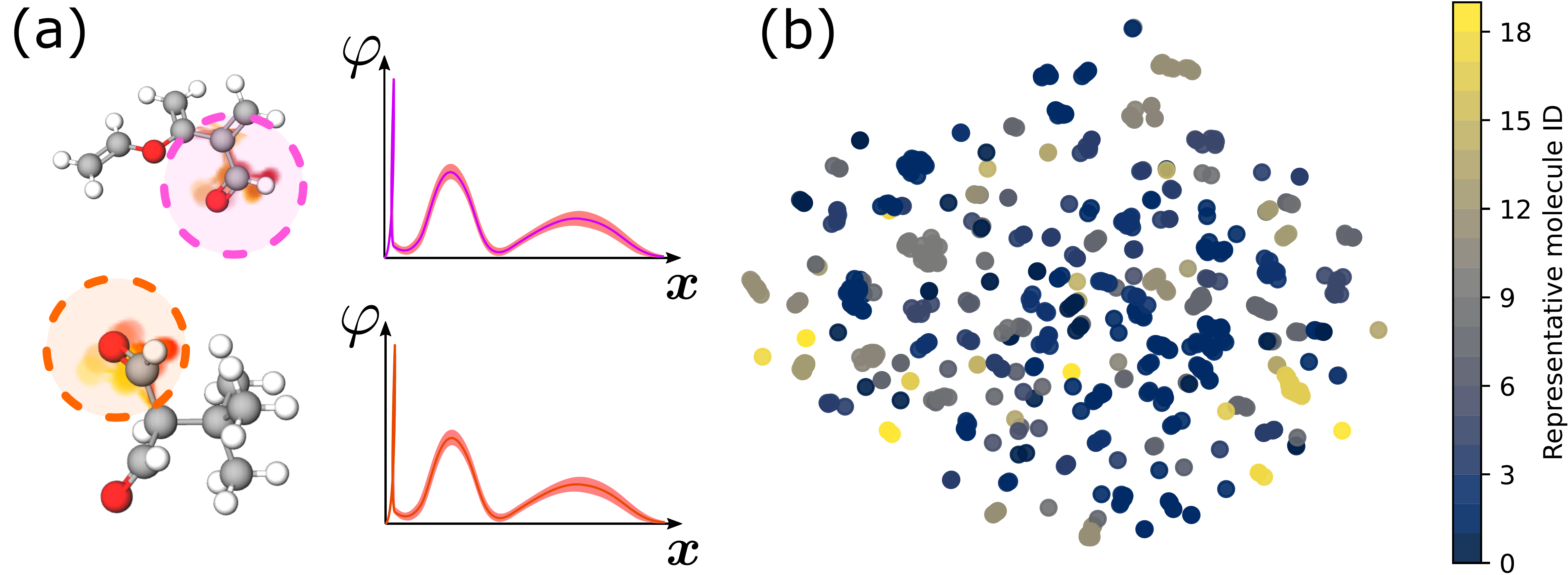}
    \caption{(a) Atomic environments averaged over gas-phase simulations are
      encoded in aSLATM vectors, $\varphi({\bm x})$. (b) UMAP projection of the
      averaged aSLATM vectors for the 3,441 C\textsubscript{7}O\textsubscript{2}
      isomers. Points colored according to one of the 19 representative
      molecules.}
    \label{fig:ee_umap}
  \end{center}
\end{figure*}

All of the gas-phase aSLATM cluster centers were combined and
clustered using \textsc{hdbscan}. We used the default \textsc{hdbscan}
parameters, with both the minimum cluster size and number of nearest
neighbors set to five points. Fig.~\ref{fig:ee_umap}b shows a UMAP projection of
this data set colored by the identified representative molecules. It
clearly shows that the set of representative covers the conformational
space of all compounds. The UMAP projection (set using the default parameters) is used only for
visualization purposes, while the identification of clusters was
performed in the high-dimensional aSLATM space. Beyond the overall separation
of aSLATM vectors based on chemical element, no other global trends
are seen across the various clusters defined. Although we only provide
labels for a small fraction of the clusters identified in
Fig.~\ref{fig:ee_umap}, we saw that most of the distinct clusters that
are present in the UMAP projection are also labeled as distinct clusters
according to our \textsc{hdbscan} results on the high-dimensional
data. Because we were also able to identify the key chemical motifs
that define these clusters via visual inspection, we are confident in
the accuracy of the clustering results. We then chose representative
molecules by first ranking them by the number of clusters ``visited,'' meaning we prioritized the compounds with aSLATM vectors belonging to as many different clusters as possible.
We then included subsequent molecules if the number of new clusters
visited by the molecule was greater than the number of clusters
already visited by the other chosen molecules. By applying this simple
algorithm, we found 19 molecules containing local environments that
shared cluster assignments with over 92\% of the assigned aSLATM
vectors. These nineteen representative molecules, shown in Figs.~S1-S4, were then used as the
foundation for our extended-ensemble approach.

\subsection{Atomistic simulations of bulk liquid-phase binary mixtures}

An extended ensemble consisting of bulk liquid-phase molecular dynamics
simulations of each of the 19 representative molecules, as well as binary
mixtures of the representative molecules, was constructed. Each system consisted
of 400 molecules in total, with the concentrations for compounds in the binary
mixtures ranging from 20\% to 80\% in 20\% increments. Therefore, the total number of
state points simulated at the atomistic resolution was 703: 19 pure
liquids plus every possible combination of binary mixtures, each simulated at the four
different concentrations.

Each of these 703 systems was simulated using the following protocol, adapted from the procedure used by Dunn and Noid \cite{dunn2015bottom}. 
400 molecules were first randomly placed into an isotropic box with a volume of 1000
nm\textsuperscript{3}. The system was energy-minimized and then run in the $NVT$ 
ensemble using a velocity-rescaling thermostat for 2~ns at a temperature of
1000\,K \cite{bussi2007canonical}. The system was then cooled to 300\,K over the
course of the next 10 ns. At this point the Berendsen thermostat and barostat were
used to reduce the size of the box and equilibrate the system in the $NPT$ ensemble
at 300\,K and 1\,bar \cite{hunenberger2005thermostat}. The resulting densities
ranged from $\approx 0.80$ g/cm\textsuperscript{3} to $\approx
  1.0$~g/cm\textsuperscript{3}. While no specific density data could be obtained
for these 19 representative molecules, these densities roughly agree with those
of 1,7-heptanediol (0.95 g/cm\textsuperscript{3}),  heptanoic acid (0.92
g/cm\textsuperscript{3}), and pentyl acetate (0.87 g/cm\textsuperscript{3}),
which also consist of 7 carbon and 2 oxygen atoms \cite{kim2016pubchem}. In a
similar vein, we were unable to find previously-reported isothermal
compressibilities for these specific compounds, and used the isothermal
compressibility of heptanoic acid, $7.4 \cdot 10^{-5}$ bar\textsuperscript{-1}
for all systems \cite{vong1997densities}. Production runs were then carried out
under these conditions in the $NPT$ ensemble using the Nos\'{e}-Hoover thermostat and
the Parinello-Rahman barostat with coupling constants of $\tau_T = 0.5$ ps and
$\tau_P = 5.0$ ps, respectively \cite{dunn2015bottom}. The force field
parameters used were the same as those used in the gas-phase simulations, with
LINCS constraints applied to the hydrogen-to-heavy-atom bonds. The final
trajectories consisted of 60~ns simulations of each system, of which the first
5~ns were discarded to allow for equilibration after applying the new thermostat
and barostat.

\subsection{Applying the multi-scale coarse-graining technique}

We briefly outline the MSCG method here, but refer the reader for a
more in-depth description~\cite{Izvekov:2005d, Izvekov:2005e,
Noid:2008a, Noid:2008b, noid2013perspective, Rudzinski:2015b}. The
first step in the coarse-graining process is to define a mapping
function from atoms to CG beads \cite{noid2013perspective}. The loss
of resolution makes the mapping choice an important one, although, in
practice, this is often based on chemical intuition alone. The
analysis of the clusters shown in Fig.~\ref{fig:ee_umap} naturally
points to a mapping scheme corresponding to functional groups. As a
result, we adopted a mapping scheme in which all combinations of two-
and three-heavy-atom fragments consisting of carbon and oxygen are
assigned to different bead types, as shown in
Table~\ref{table:ee_beadtypes}. In order to ensure completeness of our
training set---all heavy atoms are assigned to a bead type and the
topology of the fragments are preserved---we also included two
fully-branched bead types mapping to four-heavy-atom fragments. This
set of bead types led to mapping degeneracy for a number of molecules,
i.e., they can be mapped in multiple ways. The full set of compounds
and associated CG mappings is shown in Fig.~S1. Although the cartoon
mappings shown in these figures in some cases depict the beads as
being ellipsoidal, the potentials assigned to each bead type are
radially symmetric (corresponding to a spherical shape). 

\begin{table}[htb]
  \begin{center}
    \begin{tabular}{| c | c || c | c || c | c |}
      \hline
      CG Type & Fragment     & CG Type & Fragment      & CG Type & Fragment          \\
      \hline
      B01     & \texttt{CC}  & B06     & \texttt{CCO}  & B11     & \texttt{C=CO}     \\
      B02     & \texttt{CO}  & B07     & \texttt{COC}  & B12     & \texttt{OC=O}     \\
      B03     & \texttt{C=C} & B08     & \texttt{OCO}  & B13     & \texttt{C(C)(C)C} \\
      B04     & \texttt{C=O} & B09     & \texttt{CC=C} & B14     & \texttt{C(C)(C)O} \\
      B05     & \texttt{CCC} & B10     & \texttt{CC=O} &         &                   \\

      \hline
    \end{tabular}
    \caption{Bead types and their corresponding fragments in SMILES notation.}
    \label{table:ee_beadtypes}
  \end{center}
\end{table}

We now turn to determining the CG potential. In order to maintain
thermodynamic consistency condition across both CG and atomistic
systems, the marginal probabilities over the CG degrees of freedom
between the CG model and reference atomistic simulations must be
equal.\cite{Noid:2008a, noid2013perspective} Under this condition,
solving for the CG force field yields a projection of the atomistic
free-energy surface onto the CG degrees of freedom, known as the MBPMF
\cite{noid2013perspective}. We use the MSCG approach to variationally
determine a CG potential that best approximates the MBPMF
\cite{izvekov2005multiscale}. The variational principle ensures that
the resulting CG potential best reproduces the averaged atomistic net
force acting on CG sites. For this reason, the MSCG approach is also
commonly referred to as the force-matching method for bottom-up
coarse-graining. The high-dimensional MBPMF is often projected onto
molecular mechanics terms commonly used in atomistic MD, including
nonbonded pairwise contributions. Due to the inherently many-body
nature of the MBPMF, the use of pairwise interactions in the CG force
field, while computationally convenient, usually introduces some
degree of error due to the projection of many-body effects onto a
pairwise basis. However, in this work, we limit ourselves to pairwise
non-bonded interactions between the different bead types, represented
using a set of flexible spline functions as a basis set.
If the CG forces depend linearly on the parameters of the model, $\bphi$, then
the MSCG method corresponds to a linear least-squares problem in these parameters. This optimization problem can equivalently be expressed as a coupled set of linear equations (i.e., the normal equations):
\begin{equation}
  \label{eq:ee_mscg7}
  \sum_{D'}G_{DD'}\phi_{D'} = b_D,
\end{equation}
where $D$ denotes a single interaction type at a specified distance.
In this equation, the correlation matrix, $G_{DD'}$, measures the
cross-correlations between all atomistic interactions when projected
onto the force-field basis vectors defined. $b_D$ is a vector obtained
by projecting the MBPMF of the atomistic reference onto these force
field basis vectors. Solving equation \ref{eq:ee_mscg7} yields the
parameters $\phi_{D'}$ corresponding to the CG potential that
minimizes the force-matching functional.

We used the \textsc{bocs} software package developed by Dunn \emph{et al.}~to
apply the MSCG method to each of the 703 atomistic ensembles in the extended ensemble
\cite{dunn2017bocs}. For systems made up of compounds with multiple mappings, we
systematically applied every possible mapping (or combination of mappings in the
case of binary mixtures) and calculated the MSCG potential from each mapped
atomistic trajectory. We first applied the direct Boltzmann inversion method in
order to obtain intramolecular (i.e., ``bonded'') CG potentials. 
In cases where certain angle and dihedral values were not sampled,
we modified the
resulting potential to include large barriers, effectively preventing the CG
systems from sampling these values. 
To properly account for the contribution of these intramolecular interactions to the mean force,
we explicitly calculated the contributions 
and subtracted them before solving equation~\ref{eq:ee_mscg7}~\cite{Mullinax:2010a}, including only the
nonbonded and bond interactions. 
Although the bond interactions are included in the calculation, we do not update the corresponding forces (i.e., the Boltzmann-inverted bond potentials are used for all simulations).
Previous work has suggested that the inclusion of the bond interactions, even after subtracting their contribution to the mean force, can provide numerical stability for determining optimal nonbonded parameters~\cite{Rudzinski:2014,Rudzinski:2014b,Rudzinski:2021}.
All pairwise interactions were represented with radially-isotropic fourth-order basis splines with control
points spaced every 0.01 nm ranging from 0.0 to 1.4 nm. In this fashion, a set of CG
pairwise potentials was generated for each mapping at each state point. This
protocol was applied using an automated framework, and, to the best of our
knowledge, this is the first study in which such a large number of systems has
been systematically coarse-grained using the MSCG method.

\subsection{Averaging over the extended ensemble}

Mullinax and Noid proposed the extended-ensemble MSCG framework, which
extends the variational principal of the MSCG method to determine the
optimal approximation to a generalized MBPMF, constructed from a
number of system-dependent MBPMFs~\cite{mullinax2009extended}. Within
the extended ensemble, the average of an observable, $\langle A
\rangle$, is evaluated as
\begin{equation}
  \label{eq:ee_1}
  \langle A_\Gamma (\bm R_\Gamma) \rangle = \sum_\Gamma^{N_\Gamma} P_\Gamma \langle A_\Gamma (\bm R_\Gamma) \rangle_\Gamma,
\end{equation}
where $\Gamma$ specifies the molecular identity, CG mapping, and
thermodynamic state point of a single system within the extended
ensemble (i.e., a mapped ensemble as previously defined), ${\bm
R_\Gamma}$ represents the cartesian coordinates of system $\Gamma$,
and $N_{\Gamma}$ is the total number of systems making up the extended
ensemble. $P_\Gamma$ is the weight of system $\Gamma$, and is taken to
be $1/N_{\Gamma}$ in this work. $\langle \cdot \rangle_\Gamma$ denotes the
usual ensemble average within system $\Gamma$, and implies the
appropriate conditional averaging for observables evaluated from
atomically-detailed simulations. Similarly to the original MSCG
framework, the optimal CG force field parameters, $\bphi$, within the
extended ensemble can be determined by solving
Equation~\ref{eq:ee_mscg7}, while evaluating the correlation functions
according to Equation~\ref{eq:ee_1}~\cite{mullinax2009extended,
dunn2017bocs}.
 
In practice, we first initialize a  correlation matrix $G_{DD'}$ and mean force vector $b_D$
for all 105 pairwise interactions between the 14 bead types that we
have defined as well as all bonded interactions (to ensure numerical
stability). With all elements initially set to zero, we then iterate over all of the mapped ensembles,
adding each of the blocks of the correlation matrix and segments of
the mean force vector for a single state point to the corresponding
block and segment in the extended ensemble correlation matrix and mean
force vector, respectively. As multiple mappings can exist for a
single ensemble, we use the same atomistic trajectory multiple times
to efficiently obtain correlations. For example, Fig.~\ref{fig:avgjsdstest}b
shows that Molecule 21 has two different mappings, labeled mapping 0
and mapping 1. Although the number and type of beads does not change,
the way in which the atomistic fragments are mapped to these beads
does change. In this case, two distinct sets of pairwise interaction
statistics for the same interactions from a single atomistic
trajectory are obtained. In addition to the Molecule 21 case, Fig.~S1 shows
several different mappings that are applied to the same compound, similarly allowing for additional correlations to be included without generating additional atomistic trajectories. After inlcuding the correlations from each of these mapped ensembles to $G_{DD'}$ and $b_D$, we 
compute the average by dividing by the total number of mapped
ensembles as required by Equation~\ref{eq:ee_1}. Using the \textsc{bocs} software package,
we solved Equation~\ref{eq:ee_mscg7} with the extended ensemble
correlation matrix and mean-force vector.
\subsection{Validation and quantifying structural accuracy}

Once we have obtained our CG potentials, we compare state-point (SP)
specific CG potentials to the extended-ensemble (EE) potentials. Both
approaches share the same intramolecular potentials. The CG
simulations are run in the $NVT$ ensemble using an isotropic box that
has dimensions matching the average density calculated from the
atomistic state-point trajectory. A time step of $\delta t=
0.002~\tau$ was used for all simulations, where $\tau$ is the natural
time unit for the propagation of the model defined in terms of the
units of energy $\mathcal{E}=1$\,kJ/mol, mass $\mathcal{M}=1$\,amu,
and length $\mathcal{L}=1$\,nm, as $\tau =
\mathcal{L}\sqrt{\mathcal{M}/\mathcal{E}}$. The simulations were run
for $5\times 10^6$ time steps, with every 500\textsuperscript{th}
frame saved as output, and the first 500 output frames were discarded.
The \textsc{gromacs} 5.1 package was used to run all CG simulations in
this work \cite{abraham2014gromacs}. We observed a speed-up factor of
$\approx 3.0$ when comparing the CG to the atomistic simulations (with
the CG simulations running at $\approx 0.35$~ns/CPU hour).

To assess the effectiveness of the EE potentials, we first compare
radial distribution functions (RDFs), $g(r)$, between the different
models. We quantify the agreement between the CG and atomistic RDFs
using the Jensen-Shannon divergence (JSD) \cite{lin1991divergence}.
Divergences relating two functions have successfully been used in the
context of the relative-entropy framework as a useful tool for
evaluating the quality of CG models \cite{chaimovich2011coarse,
foley2015impact}. We previously used the JSD to evaluate the CG
distribution of water/octanol partitioning free energies across small
organic molecules,\cite{kanekal2019resolution} as well as force-field
accuracy within the conformational surface hopping
scheme.\cite{rudzinski2020coarse} While the Kullback-Leibler
divergence, $D_{\mathrm{KL}}$, \cite{Kullback1951} directly relates
two distributions, the JSD computes the relative entropy by comparing
each of these distributions to the average of the other two
\begin{equation}
  \label{eq:ee_jsd}
  D_{\mathrm{JS}} = \frac{1}{2}D_{\mathrm{KL}}\left(g_{\mathrm{CG}}(r)||g_{\mathrm{avg}(r)}\right)
  + \frac{1}{2}D_{\mathrm{KL}}\left(g_{\mathrm{AA}}(r)||g_{\mathrm{avg}}(r)\right),
\end{equation}
where
\begin{align*}
  D_{\mathrm{KL}}(g_{\mathrm{A}}(r)||g_{\mathrm{B}}(r)) & = \sum_{r=0}^{r_{\rm max}} g_{\mathrm{A}}(r) \ln\left(\frac{g_{\mathrm{A}}(r)}{g_{\mathrm{B}}(r)}\right), \\
  g_{\mathrm{avg}}(r)                                   & = \frac{1}{2}(g_{\mathrm{CG}}(r)+g_{\mathrm{AA}}(r)).
\end{align*}
In the above equations,
we define $D_{\mathrm{KL}}$ in terms of two arbitrary RDFs, $g_{\mathrm{A}}(r)$
and $g_{\mathrm{B}}(r)$ ranging from $r=0$ to $r_{\rm max}$.
For all RDFs, we used a
grid spacing of 0.01 nm and $r_{\rm max} = 1.5$~nm. All RDFs were calculated
using the \textsc{gmx rdf} package included in \textsc{gromacs} 5.1. The JSDs
for both the SP and EE are compared to their respective atomistic RDFs.

\subsection{Mean Force Decomposition Analysis}
\label{sec:MFD}

Equation~\ref{eq:ee_mscg7} can be transformed to depend only on structural information, revealing the set of equations as a generalization of the Yvon-Born-Green integral equation framework from liquid state theory~\cite{Mullinax:2009b,Mullinax:2010}.
Within this formulation, for a single pairwise-additive distance-dependent interaction represented with a set of piecewise constant basis functions, $\bb$ corresponds to a structural correlation function that is directly related to the radial distribution function (RDF):
\begin{equation}
  b_D = k_{\rm B} T c R_D^2 \left (\frac{{\rm d} {\bm g}}{{\rm d} R} \right )_{\!\!D} \,\, ,
  \label{eq:bd}
\end{equation}
where $c=(4\pi N)/(3 V)$ and $\bm g$ is the discretization of the RDF implied by the basis function representation.
$\left ({\rm d} {\bm g} / {\rm d} R \right )$ is meant as a numerical derivative of $\bm g$ with respect to interparticle distance $R$, given by the basis function centers $\{R_D\}$.

The correlation matrix $\bG$ also has a clear physical interpretation~\cite{Rudzinski:2012vn}.
First, it is useful to decompose $\bG$ into two matrices which, through Equation~\ref{eq:ee_mscg7}, determine the direct and indirect contributions to $\bb$:
\begin{equation}
  G_{DD'}=\bar{g}_D \delta_{DD'} + \bar{G}_{DD'} \,\, ,
  \label{eq:decompose}
\end{equation}
where $\delta_{DD'}$ is the Kronecker delta function.
The direct contribution $\bar{\bm g}$ is a correlation function that is again related to the RDF: $\bar{g}_D = c R_D g_D$.
$\bar{\bG}$, on the other hand, quantifies the cross correlations between pairs of interactions, in this case the average cosine of the angle formed between triplets of CG sites~\cite{Rudzinski:2012vn}.
%
Equation~\ref{eq:bd} clearly implies a relationship between $b(R)$ and the pair mean force, $-w'(R) = -\frac{\rm d}{{\rm d}R} [-\kT \ln  g(R) ]$.
Thus, using Equation~\ref{eq:decompose}, the pair mean force can be decomposed into direct and indirect contributions:
\begin{equation}
  -w'_D = \frac{b_D}{\bar{g}_D} = \phi_D + \frac{1}{\bar{g}_D}  \sum_{D'} \bar{G}_{DD'} \phi_{D'} \,\, .
  \label{eq:ppmf}
\end{equation}

\section{Results}

In this work, we construct a chemically-transferable and structurally-accurate CG model for C\textsubscript{7}O\textsubscript{2} isomers, following a bottom-up approach.
The model was parametrized using an ``extended ensemble'' of 703 atomistic reference ensembles of pure liquids and binary mixtures, consisting of 19 representative compounds determined by clustering the gas-phase conformation-averaged atomic SLATM vectors of 3,441 C\textsubscript{7}O\textsubscript{2} isomers. 
The parametrization also included multiple CG mappings for individual reference systems, resulting in 2,476 mapped ensembles in total (see Fig.~S1). 

\begin{figure}[htbp]
  \begin{center}
    \includegraphics[width=\linewidth]{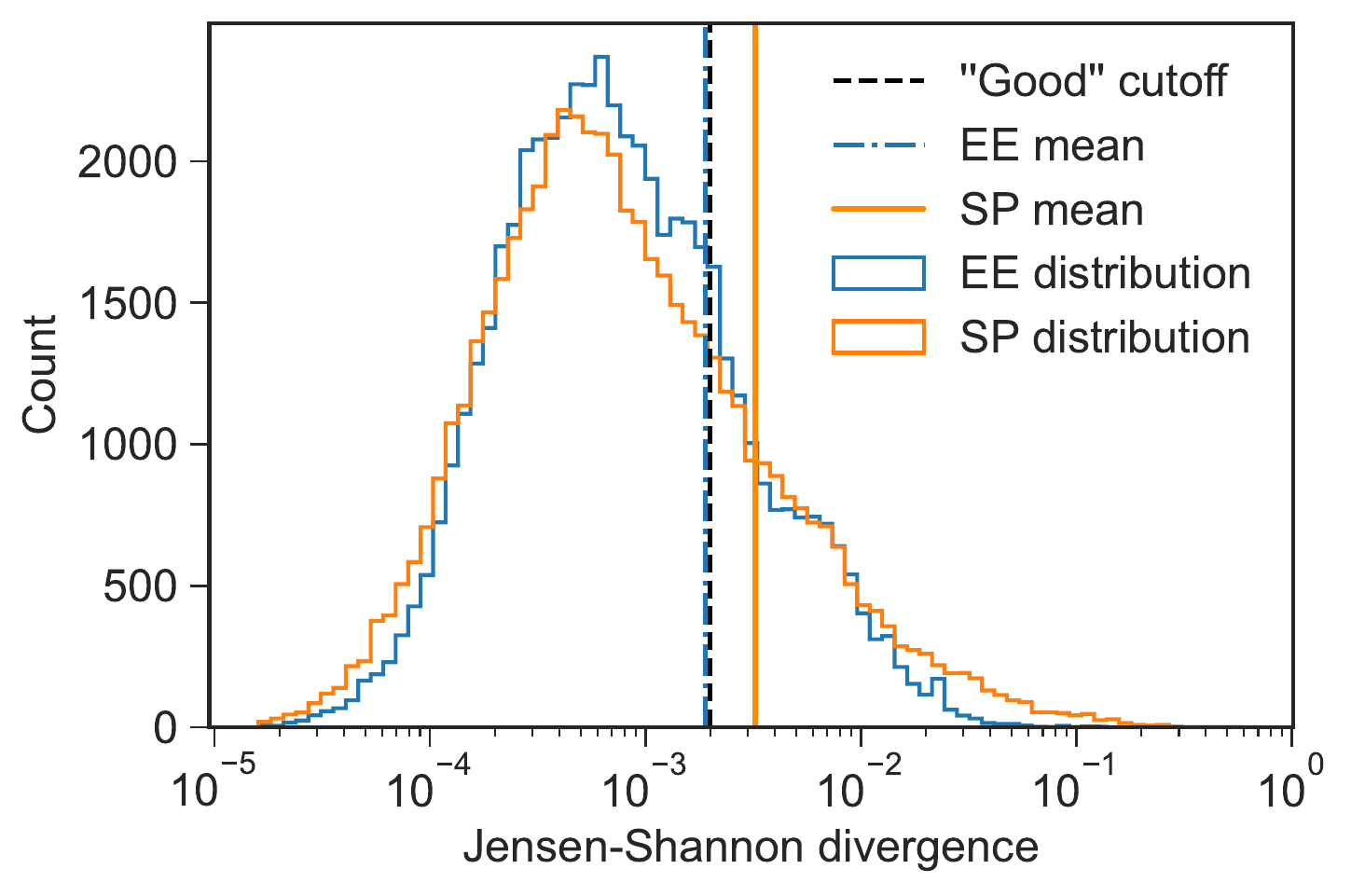}
    \caption{Distribution of JSD values using both the state-point
      specific (orange) and transferable extended-ensemble (EE; blue)
      models. The black dashed line denotes the cutoff JSD value for
      ``good'' agreement with atomistic RDFs, 0.002.}
    \label{fig:avgjsdspermapping}
  \end{center}
\end{figure}

In the following, the extend ensemble (EE) model is assessed through
comparisons of RDFs to both the reference atomistic ensembles (at the
CG level of resolution) and also to state-point specific (SP) models,
i.e., models constructed using individual reference simulations. SP
and EE parametrizations share all intramolecular (i.e., bond, angle,
and dihedral) interactions, obtained by direct Boltzmann inversion of
the pure-liquid simulations. Each of the 2,476 mapped ensembles
contains up to 28 RDFs, making a manual inspection unfeasible
(although all EE RDFs are available online~\cite{zenodo}). Note that while the atomistic simulations were run in the
$NPT$ ensemble, the CG simulations were run in the $NVT$ ensemble,
with the volume of the simulation box equal to the average volume of
the atomistic simulation box. We assess the relative error of the CG
models at a density corresponding to the atomistic reference. We use
JSD values to quantify the accuracy of the SP and EE CG RDFs relative
to the atomistic RDFs. Fig.~S2 provides several
examples of RDF comparisons that result in certain JSD values, a
useful reference for interpreting these JSD values in terms of the
error when comparing atomistic and CG RDFs.
Fig.~\ref{fig:avgjsdspermapping} reports the distribution of JSD
values for all 36 homogeneous liquids (see Fig.~S3 for the state-point averaged JSDs per system). Also shown in this
figure are the mean of both the SP and EE CG models. One might expect
the EE model to perform worse than the SP models because the EE model
is obtained by averaging over many different reference ensembles,
rather than optimizing the model for any particular one. Remarkably,
on average, the transferable EE model outperforms the SP models with
an average JSD value of 0.0024 versus 0.0038, respectively. The EE
distribution is also \emph{narrower}, indicating more regularity in
the quality of the CG parametrizations. We find several state points
where the EE model greatly outperforms the SP model: Molecule 3
mapping 0, Molecule 8 mapping 0, Molecule 1 mapping 0, see
Fig.~S3. On the other hand, we also find
opposite cases: Molecule 6 mapping 0 and Molecule 5.

\begin{figure*}[htbp]
  \begin{center}
    \includegraphics[width=0.8\linewidth]{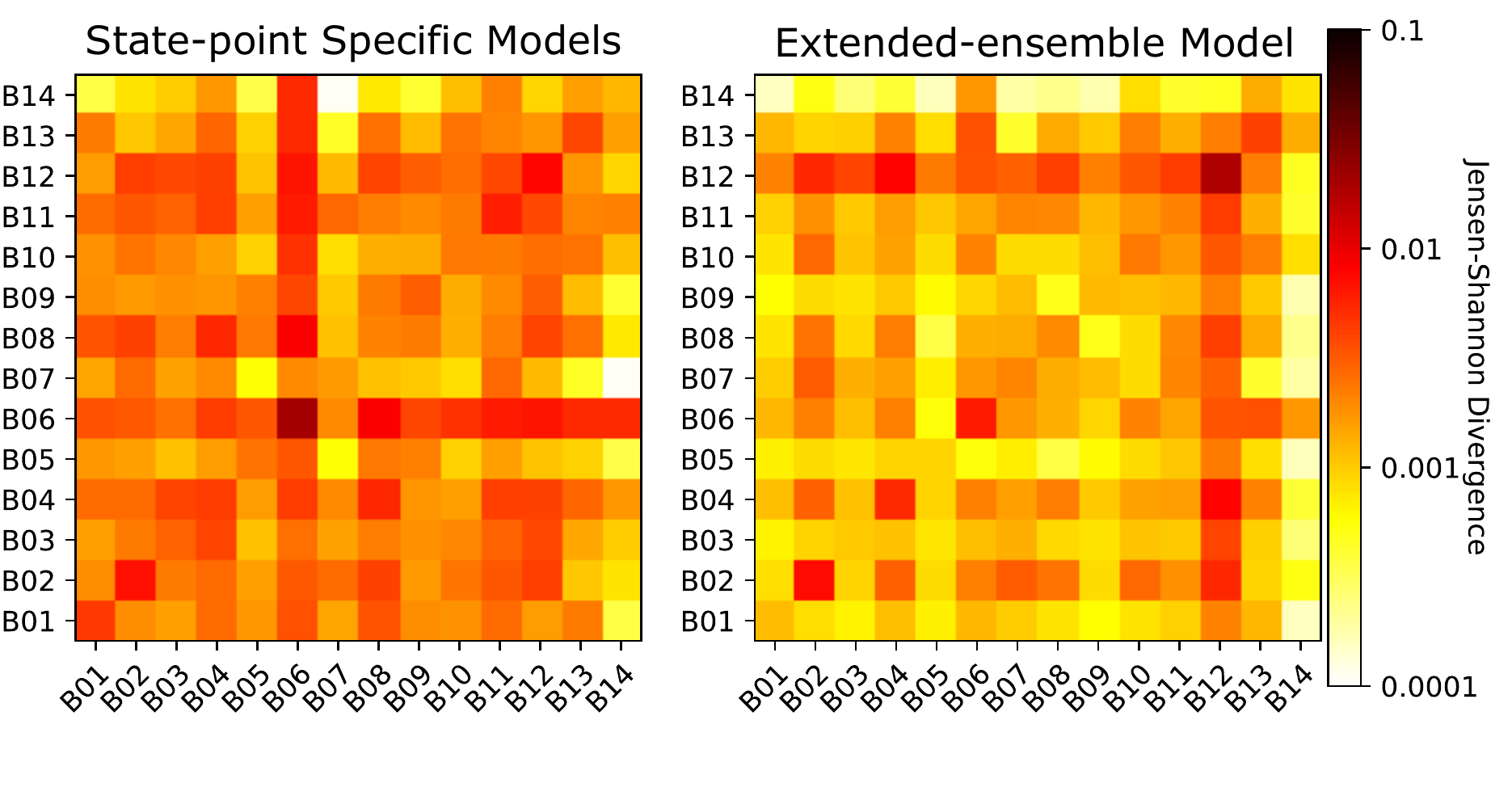}
    \caption{JSD values of interactions sampled in pure, homogeneous
      liquids using both the state-point (SP) specific and
      transferable extended-ensemble (EE) CG models on the left and
      right panels, respectively.}
    \label{fig:avgjsdsperinter}
  \end{center}
\end{figure*}

We now change perspective: we analyze the same set of systems and
RDFs, but average according to \emph{interaction types}.
Fig.~\ref{fig:avgjsdsperinter} presents a matrix-form heat map of JSD
values, with column-row combinations representing interaction pairs.
The lighter coloring of the EE interactions conveys the same message
as before: EE CG models are on average closer to the atomistic
reference, and the SP CG models show more outliers. The use of a
logarithmic scale emphasizes strong deviations. While most of the EE
RDFs are significantly below the ``good'' agreement JSD cutoff, the
previous averaging over systems leads to larger JSD values (Fig.~S3).
The difference in the tails of the SP and EE distributions in
Fig.~\ref{fig:avgjsdspermapping} highlights the dominating effect of a
few interaction types. 

We now investigate the transferability of the EE model beyond the set
of representative molecules, but within the considered chemical space
of 3,441 C\textsubscript{7}O\textsubscript{2} isomers. ``Test''
compounds are selected based on their molecular SLATM distance from
the training compounds. The molecular SLATM vector simply consists of
the sum of aSLATM vectors in a molecule. We quantify compound
similarity from the 3,441 isomers to the 19 representative molecules
by means of a matrix of pairwise Euclidean distances between molecular
SLATM representations. To focus on molecules that share as little
information as possible from the pool of representative molecules, we
focus on the largest \emph{average} distances.
Tab.~\ref{table:ee_testset} reports the SMILES string of the five
furthest compounds, as well as their scaled SLATM distance (i.e., the
maximum Euclidean distance is 1.0).

\begin{table}[htbp]
  \begin{center}
    \begin{tabular}{| c | c | c | c | c | c |}
      \hline
      Molecule & SMILES string               & Scaled SLATM distance \\
      Index    &                             & from training set     \\
      \hline
      19       & \texttt{CCC(CC)OC(C)=O}     & 0.43                  \\
      20       & \texttt{CC(C)=CC(=C)C(O)=O} & 0.48                  \\
      21       & \texttt{C=COC(=C)C(=C)C=O}  & 0.88                  \\
      22       & \texttt{CC(C)(C)C(C=O)C=O}  & 0.91                  \\
      23       & \texttt{CC(C)C(C)(C=O)C=O}  & 0.91                  \\
      \hline
    \end{tabular}
    \caption{Test molecules, SMILES strings, and SLATM distance to the
      representative molecules scaled by the maximum distance.}
    \label{table:ee_testset}
  \end{center}
\end{table}

\begin{figure}[htbp]
  \begin{center}
    \includegraphics[width=\linewidth]{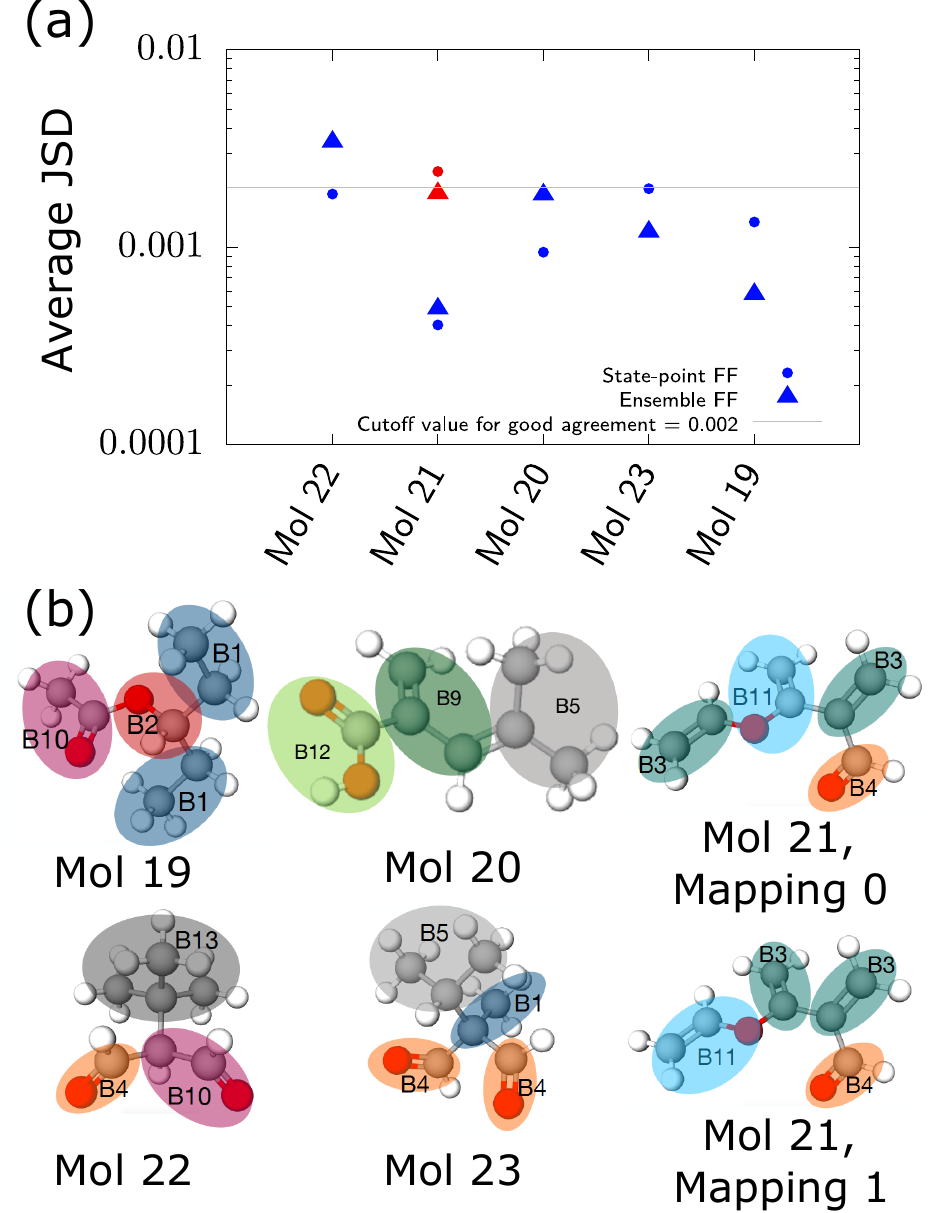}
    \caption{Average JSD values of bulk liquid MD simulations for five
      test compounds, displaying both SP (circles) and EE (triangles)
      CG models. Molecule 21 has two mappings, shown in different
      colors. The grey line denotes the cutoff JSD value for ``good''
      agreement with atomistic RDFs, 0.002. The molecules are ordered
      based on increasing agreement of the atomistic RDFs with the EE
      CG RDFs. }
    \label{fig:avgjsdstest}
  \end{center}
\end{figure}

The performance of the CG models for the test molecules, as well as an
illustration of their mappings, is shown in
Fig.~\ref{fig:avgjsdstest}. In analogy to
Fig.~\ref{fig:avgjsdspermapping}, we average the JSDs of the SP and EE
CG RDFs for each system. We find that the largest improvement from SP
to EE parametrization corresponds to Molecule 19---the closest
compound to the representative set. It confirms that a larger
conformational overlap can benefit the transferable-parametrization
strategy. Other factors also play a role, as indicated by the superior
and comparable performance of the EE model for Molecules 23 and 21,
respectively, despite these molecules being further away from the
representative set on average (see Table~\ref{table:ee_testset}). On
the other hand, the EE model underperforms compared to the SP model
for Molecules 20 and 22. While Molecule 22 is also one of the furthest
compounds on average from the representative set, Molecule 20 is only
slightly further than Molecule 19. We defer a rationalization of the
results for these compounds to later in the text. Evidently, an
analysis of five molecules is by no means statistically representative
of the chemical space considered. However, this provides a glimpse of
the behavior of the EE model for molecules with varying conformational
overlap.

\section{Discussion}

Our results show that an extended-ensemble (EE) parametrization across
a wide set of small organic isomers leads to more accurate and
consistent CG models. This was demonstrated in
Fig.~\ref{fig:avgjsdspermapping}, where the distribution of EE JSD
values shows a smaller mean and variance than the state-point specific
(SP) models. These results might be counterintuitive, in that a force
field that is parametrized using information averaged over many
simulations is expected to perform worse than another of equal
complexity that focuses on a particular reference ensemble. Instead
the results indicate that better transferability can go hand in hand
with improved accuracy. Beyond this overall improved accuracy, the
reduced variance of the JSD distribution indicates that the EE model
will result in more reliable predictions. On the other hand, our
analysis also reveals cases where the EE model underperforms, compared
to a more traditional SP parametrization. To better understand the
advantages and pitfalls of the EE parametrization, we investigate certain 
ensembles and the corresponding mapped ensembles where the EE and SP models lead to significant differences.

\begin{figure*}[htbp]
  \begin{center}
    \includegraphics[width=0.898\linewidth]{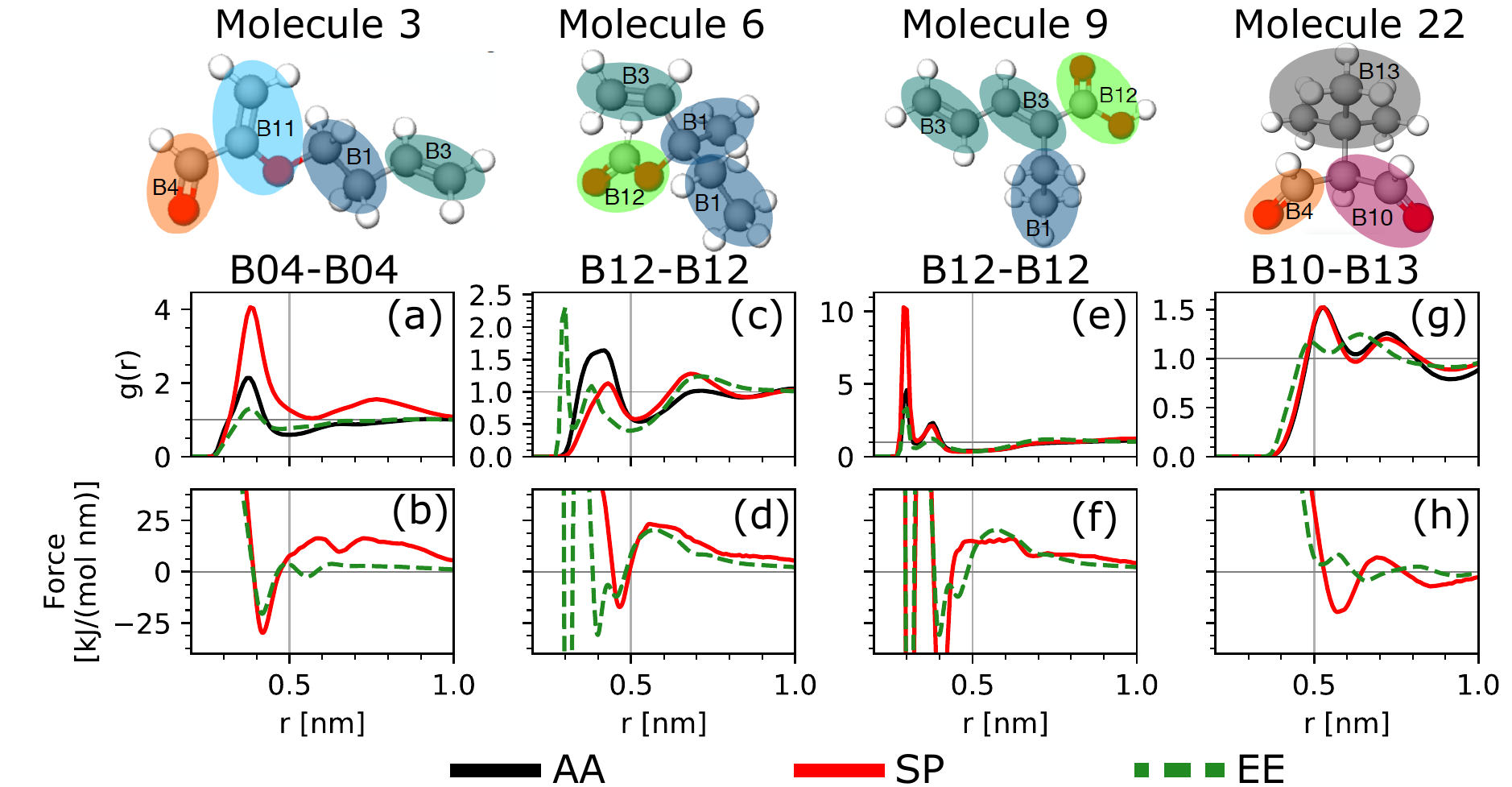}
    \caption{Atomistic and CG RDFs of pure-liquid simulations and
    pairwise forces for molecules (a-b) 3, (c-d) 6, (e-f) 9, and (g-h)
    22, respectively. The black, red, and green curves denote
    respectively the atomistic, SP, and EE RDFs for the fragments
    which map to the bead types listed in the top-right of each plot.}
    \label{fig:mol_rdfs}
 \end{center}
\end{figure*}

We first consider the pure Molecule-3 system. Mapping 0, depicted in
the molecular image at the top of Fig.~\ref{fig:mol_rdfs}, shows the
greatest structural improvement from SP to EE parametrization,
according to the average JSD value (Fig.~S3).
An example RDF for the B04--B04 interaction is shown in
Fig.~\ref{fig:mol_rdfs}a, and the RDFs pertaining to all other
pairwise interactions are available in the SI (See Fig.~S5). The SP model
(solid red curve) drastically overstabilizes the first and second
solvation peaks of the B04--B04 RDF, while the EE model (dashed green
curve) better reproduces the AA simulation, with a mild
understabilization of the solvation structure.
Fig.~\ref{fig:mol_rdfs}b presents the B04--B04 pair forces for the SP
and EE models. Both forces exhibit similar features within the first
solvation shell region, with minima at $r \approx 0.4$\,nm. However,
the EE force demonstrates a significant reduction of the magnitude of
repulsive forces beyond this minimum. Overall we found that the magnitude of these repulsive features were always either maintained or reduced in the EE model with respect to the SP model, and were rarely seen to increase in the EE case. 
The repulsive nature of the SP forces
is consistent with previous work showing that structure-based CG
approaches tend to result in models with overly repulsive
potentials.\cite{wang2009comparative,
guenza2015thermodynamic,dunn2015bottom,rudzinski2020coarse}. Compared
to the SP models, the EE forces tend to look simpler---qualitatively
more similar to a Lennard-Jones form. A similar finding was reached
when augmenting a CG model with multiple, conformationally dependent
force fields \cite{rudzinski2020coarse}. The results suggest that
solving Equation~\ref{eq:ee_mscg7} over the extended ensemble promotes
a \emph{regularization} effect, which accounts for correlations across
conformational and chemical space. Averaging over these correlations
appears to have the net effect of smoothening sharp, localized
features in the mean force while preserving the key features shared
across the extended ensemble. 

Next, we examine cases where the EE model underperforms compared to
the SP model. Fig.~\ref{fig:avgjsdsperinter} shows that the EE
B12--B12 interaction, found in molecules 6, 9, and 16 (see SI Fig.~S5), is
significantly worse when compared to the SP model, with average JSD
values of 0.018 and 0.008, respectively. Panels (c) and (d) of
Fig.~\ref{fig:mol_rdfs} present the SP and EE B12--B12 RDFs and
forces, respectively, for molecule 6. Panels (e) and (f) show the
corresponding quantities for molecule 9. In contrast to the B04--B04
interactions of molecule 3, the repulsive bumps at $r \approx 0.6$\,nm
are retained within the EE model, suggesting that they are essential
for stabilizing the proper structure. Both the SP and EE forces for
molecule 9 contain a sharp attractive feature at $r \approx 0.3$\,nm,
which are clearly responsible for the corresponding sharp crystalline
peaks in the RDFs at this distance. A similar feature is found in the
EE force for molecule 6, although there is no corresponding feature
for the SP model. We conclude that the extended-ensemble averaging
``transferred'' this particular trait from molecule 9 to
others---including molecule 6, resulting in significant errors in the
RDFs for this interaction type. Overall, we find that within the EE
approach interactions involving B12 average over significantly
different local environments. Indeed, the atomistic RDFs of molecules
6 and 9 display pronounced differences. The sharp peaks observed in
molecule 9 clearly indicate liquid-crystalline behavior, absent in the
molecule 6 liquid. This is expected: molecule 9 consists of
alternating single and double bonds facilitating $\pi$-stacking
interactions, a clear promoter of liquid crystals for organic small
molecules \cite{kato2006functional, jackson2013controlling}.
Furthermore, the presence of a terminal carboxylic-acid group
encourages hydrogen bonding in the bulk-liquid phase, which also
promotes ordering. On the other hand, the B12 bead in molecules 6 and
16 represents ester groups, which lack hydrogen bonding, and also lack
conjugated bonds for $\pi$-stacking. Our use of a single bead type to
represent such different chemical environments results in a CG
potential that cannot faithfully reproduce either case. Furthermore,
the strong anisotropic character of $\pi$-stacking and
hydrogen-bonding interactions may call for potentials that go beyond
pairwise and isotropic functions.\cite{stillinger1983inherent,
greco2019generic} We foresee that improvements in the CG mapping
and/or the CG force-field complexity would help remedy the situation. While the reuse of atomistic trajectories to generate multiple mapped ensembles allows for an efficient way to obtain correlations and forces to average over using the EE approach, failing to account for these differences in atomic environments can lead to an exacerbation, rather than a reduction, of undesirable features in the resulting forces.

\begin{figure*}[htbp]
  \begin{center}
    \includegraphics[width=0.9\linewidth]{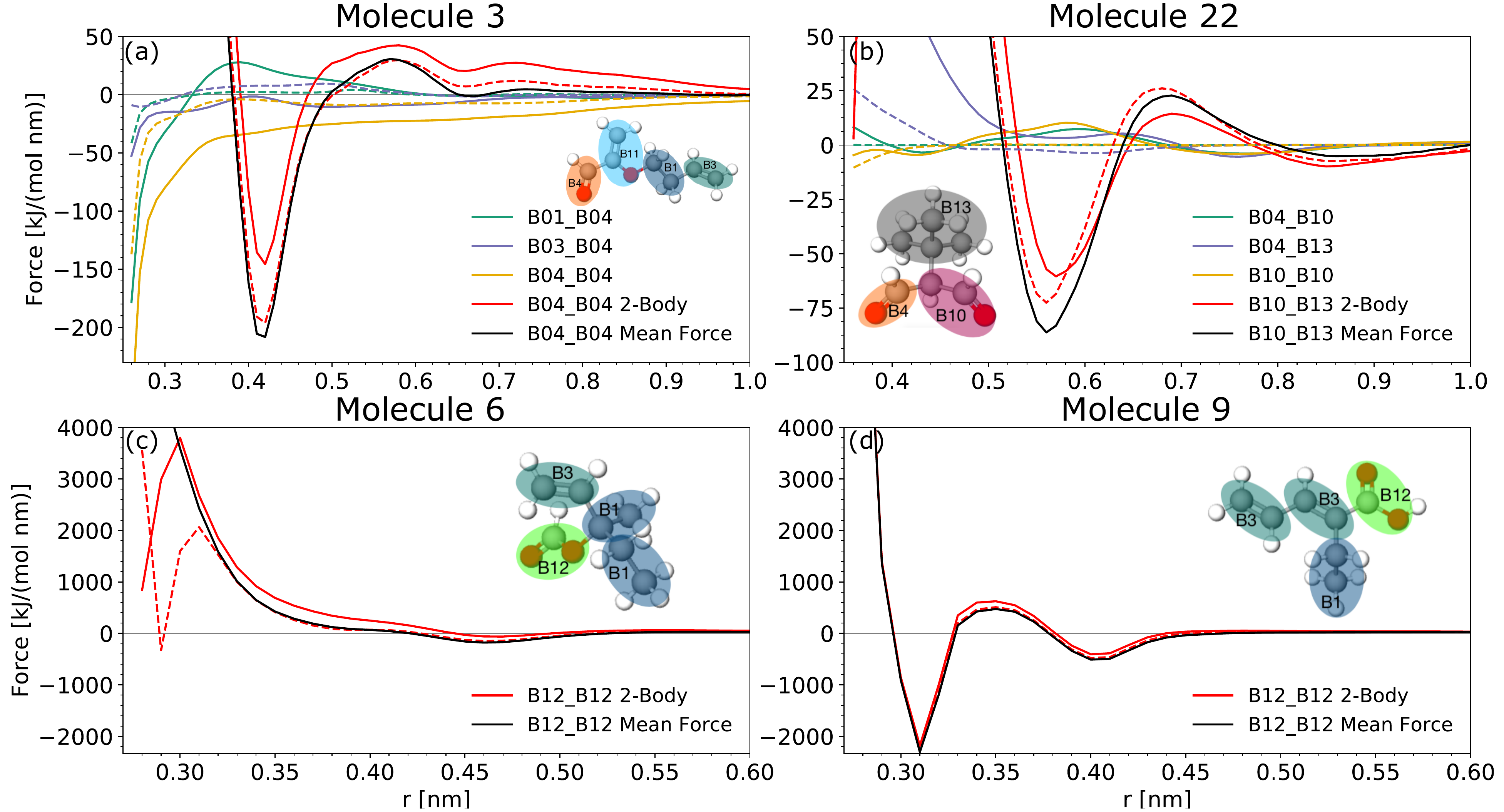}
    \caption{Mean
    forces (black curves) for the interactions corresponding to the
    RDFs shown in Fig.~\ref{fig:mol_rdfs}. For (a) and (b), the three
    of the three-body contributions to the mean force for both the SP
    (solid) and EE (dashed) models are shown. For (c) and (d), The
    two-body contributions to the B12-B12 mean force for the SP
    (solid) and EE (dashed) models are shown. }
    \label{fig:meanforcedecomp}
  \end{center}
\end{figure*}

To further understand the apparent regularization effect that arises
due to averaging correlations within the extended ensemble, we
performed an analysis of the mean forces for the pure liquid systems
of the four molecules presented in Fig.~\ref{fig:mol_rdfs}
(representative Molecules 3, 6, and 9, and test Molecule 22).
Following Section~\ref{sec:MFD}, we first decomposed the SP mean
forces into contributions from each of the interactions in the system,
using the cross-correlations calculated from the corresponding
reference ensemble. The solid curves in Panel (a) of
Fig.~\ref{fig:meanforcedecomp} present the resulting decomposition for
the B04--B04 interaction of molecule 3, for a subset of the
contributions. The remaining contributions have negligible impact on
the B04--B04 mean force (solid black curve). The solid red curve represents the direct, or 2-body,
contribution (i.e., the SP B04--B04 pair force). The other colored
solid curves represent 3-body contributions (i.e., correlated
contributions to the B04--B04 mean force from a particular distinct
interaction). By definition, the sum of 2- and 3-body contributions
equal the total mean force (Eq.~\ref{eq:ppmf}). Thus, in this case, it
is apparent that, while the 2-body contribution is dominant, there are
significant contributions from other interactions, both within the
first solvation shell and beyond. Panel (b) of
Fig.~\ref{fig:meanforcedecomp} presents the corresponding result for
the B10--B13 interaction of molecule 22, with similar overall features
to the B04--B04 case.

To directly probe the impact of averaging correlations over distinct
environments, we repeated the decomposition of the SP mean forces
using EE correlations instead of the SP correlations. The results are
presented as the dashed curves in panels (a) and (b) of
Fig.~\ref{fig:meanforcedecomp}. For both molecules 3 and 22, there is
a reduction in the magnitude of the 3-body contributions to the mean
force, as might be expected due to smoothing of correlations via the
EE averaging. This can be most clearly seen in the similarity between
the 2-body contributions (dashed red curves) and the total mean force
(solid black curves). 
To interpret these results, it helps to reconsider the g-YBG
equations. Eq.~\ref{eq:ee_mscg7} represents an exact relationship
between the force field parameters $\bphi$ and the structural
correlation functions $\bb(\bphi)$ for a single state point,
determined from molecular simulations, via the cross correlations,
$\bG(\bphi)$, generated by the same model $\bphi$. In contrast, the
MS-CG method attempts to predict the force-field parameters $\bphi$
that will reproduce $\bb^{\rm AA}$, using $\bG^{\rm AA}$ as a proxy
for the cross correlations of the CG model~\cite{Rudzinski:2014,
Rudzinski:2014b}. While ideally $\bG^{\rm AA} = \bG(\bphi)$,
limitations in the CG basis set can only approximately reproduce the
AA correlations. The EE scheme populates the correlation matrix with
complementary contributions from various systems and state points.
Incorporating more reference simulations could improve the state-point
parametrization, by smoothing out correlations that are too
complicated for the CG model to reproduce. However, this numerical
experiment represents only a portion of the extended ensemble
calculation, which additionally performs an average over the various
mean forces, i.e., through the average over the $\bb^{\rm AA}$
coefficients for each system and state point. It is apparent from the
analysis in Fig.~\ref{fig:meanforcedecomp} that the smoothing of
correlations \emph{is not} responsible for the lack of repulsive
features in the EE forces beyond the first solvation minimum, as
discussed above. This implies, instead, that the smoothing of the mean
force itself is the primary cause for the removal of these features.

Panels (c) and (d) of Fig.~\ref{fig:meanforcedecomp} present a similar
analysis for molecules 6 and 9, respectively, but only show the total
mean forces (solid black curves) and the 2-body contributions (red
curves) using SP (solid) and EE (dashed) correlations. For molecule 9
(panel (d)), which exhibits the liquid crystalline peak in the
B12--B12 RDF, both the SP and EE correlations result in a 2-body
contribution with a strong inflection (i.e., a deep minimum in the
potential) at $r \approx$ 0.3\,nm. On the other hand, for molecule 6
(panel (c)), the SP model displays no such inflection. Note that the
dip in the SP force for short distances is a numerical artifact that
sometimes occurs at the end of the sampled region. In the case of the
EE correlations, the situation is less clear. There is some sort of
inflection in the force at short distances, which could be partially
due to the correlations, or could also be a numerical artifact. Since
simulation of the resulting forces does not yield crystalline peaks,
as in the full EE case, we conclude that it is primarily the
combination of mean forces within the extended ensemble that is
responsible for transfering the liquid crystalline behavior between
systems. 

Finally, we turn our attention to the test molecules used for
validation of the EE parametrization. Molecules 19, 21, and 23 show
similar or improved performance using the EE model compared to SP. On
the other hand, the EE model underperforms for molecules 20 and 22. Reminiscent
of molecules 6 and 9, the discrepancy for molecule 20 also stems from
the poor modeling of the carboxylic-acid B12 bead type. Molecule 20 is
indeed similar to molecule 9, both featuring alternating single and
double bonds, as well as a terminal carboxylic-acid group. On the
other hand, the SP parametrization of molecule 20 is significantly
more accurate than that of molecule 9. Interestingly, SI Fig.~S5 shows
that the largest difference between SP and EE models when comparing these two molecules does not stem
from B12, but instead from the B09 bead type. The SP model features a
large repulsive peak in the B09--B09 interaction, nonexistent in the EE
model. Indeed, the EE parametrization was devoid of B09 fragments
showing liquid-crystalline behavior. The superiority of SP in this
case reinforces the need for a consistent mapping of fragments,
thereby ensuring homogeneous chemical environments.

Molecule 22 also poses a challenge for the EE parametrization. While
both molecules 22 and 23 are furthest from the representative
compounds and feature similar molecular structures, the EE
parametrization under- and overperformed compared to SP, respectively
(Fig.~\ref{fig:avgjsdstest}). Both molecules are structurally similar,
branched and symmetric with respect to the two carbonyl groups.
Critically, the CG mapping for molecule 23 is symmetric, while that of
molecule 22 is \emph{asymmetric}. The carbonyl groups in molecule 22
are unevenly split into fragments of different sizes, mapping to B04
and B10 types (Fig.~\ref{fig:avgjsdstest}). Here, symmetry appears to
impact the quality of the EE parametrization. Chakraborty \emph{et
al.}~recently showed that CG-mapping symmetry has a negligible impact
on structural accuracy \cite{chakraborty2020preservation}. Asymmetry
indeed appears to be irrelevant for SP models. However, the use of
asymmetric CG mappings will hurt the transferability in the EE scheme.
To understand why, it helps to consider the g-YBG equation
(Equation~\ref{eq:ee_mscg7}). Much of the benefit of the EE strategy
revolves around the sharing of reference atomistic information, both
within the correlation matrix $G_{DD'}$ as well as the projection of
the mean force $b_D$, thereby enriching the parametrization with
information from more reference ensembles. A symmetric choice of CG
mappings acts in a similar way on $G_{DD'}$ and $b_D$, further
enhancing the beneficial impact of the EE scheme.

All in all, our results highlight favorable transferability of the EE
parametrization for a variety of compounds, with promising prospects
across our chemical space of 3,441 isomers. Once the CG bead types
have been parametrized across the EE, the procedure readily offers
structurally accurate nonbonded CG interactions for any additional
molecule: we simply decorate them with appropriate bead types.  While capable
of offering transferable CG potentials, the gas-phase-based mapping
scheme was not able to account for some of the emergent behavior
occurring in the liquid phase. For example, we did not account for specific intermolecular interactions
(e.g., hydrogen bonding or $\pi$-stacking), leading to some
discrepancies. The fact that one such ``anomalous'' compound made its
way as a representative molecule speaks for the strength of our
clustering analysis from gas-phase trajectories alone. We hypothesize that a subsequent
clustering step on liquid-phase trajectories could help overcome this
issue. Incorporating liquid-phase simulations could help
reveal variations of local environments for the same fragment, and
could be used to optimize the number and set of CG bead types, as well
as the complexity of the CG force field.

\section{Conclusions}
We present an approach to construct chemically-transferable
coarse-grained (CG) models that preserve the liquid-phase structure of
small organic molecules. Our strategy couples unsupervised learning
methods with rigorous structure-based coarse-graining techniques.
Instead of focusing on a specific compound, we target a large
collection of molecules at once---in this study a collection of 3,441
C$_7$O$_2$ small-molecule isomers. The procedure first consists of
sampling the conformational space of each molecule, here using
gas-phase molecular dynamics simulations. We then encode the
configurational information by means of conformationally averaged
aSLATM atomic representations.\cite{Huang2020} Overlapping local
environments across the chemical space are systematically identified
using the graph-based clustering technique \textsc{hdbscan}. The
clusters are organized according to hierarchies of increasing
resolution, corresponding to the many-body types encoded in aSLATM.
Because clusters primarily differentiate on the basis of functional
groups, we choose them as our CG mapping scheme. We identify 19
representative compounds, whose local environments maximally overlap
with the rest of the chemical space. This subset of representative
compounds forms the basis of our liquid-phase simulations, both
homogeneous bulk and binary mixtures. All 703 atomistic reference
ensembles are combined to parametrize the CG potentials of our 14 bead
types, using the extended-ensemble multiscale coarse-graining
(EE-MSCG) method.\cite{mullinax2009extended} To the best of our
knowledge, no study so far has presented an EE parametrization
over such a broad chemical space.

Validation of our CG parametrization consisted of a systematic and
large-scale analysis of the structural accuracy. Radial distribution
functions are compared
between CG and atomistic resolution with in-depth analysis of certain pure (i.e., single-component) liquids that stood out as outliers. The transferability of the CG
force field is assessed by comparing the EE model to a more common
state-point specific (SP) force-field parametrization. Remarkably, we
find that the EE model outperforms the SP models, despite the EE model
being primarily parametrized from binary mixtures. Beyond the set of
representative compounds used for parametrization, validation against
five other molecules led on average to similar or better performance
with the EE model compared to the SP model.  Examination of specific
systems sheds light on the benefits of the EE approach: averaging
across the extended ensemble smoothens sharp features in the mean
force that are not shared across systems. On the other hand, key
features that persist across multiple state points are preserved.
Thus, the EE procedure effectively leads to a regularization in the
space of force fields, optimizing the force-matching functional to
more transferable solutions. However, we also found two detrimental
effects: ($i$) Averaging over significantly different chemical
environments of a given CG bead type, for instance due to strong
directional interactions, may erroneously promote liquid-crystalline
behavior for some compounds; ($ii$) An inconsistent treatment of
symmetry in CG mapping may limit the beneficial averaging effects of
the extended-ensemble approach. In these cases, averaging correlations
and mean forces over distinct reference ensembles resulted in a model
with larger structural deficiencies than the corresponding SP model.
Thankfully, there are clear avenues to remedy these aspects. EE-MSCG
parametrizations that cover broad subsets of chemical space offer an
appealing strategy toward structurally accurate high-throughput
coarse-grained modeling.\cite{bereau2020computational}

\section*{Supplementary Material}
The attached supporting information provides details on ($i$) all CG mappings used for the representative compounds; ($ii$) an alternative schematic of the methods ($iii$) a subset of the data shown in Fig.~\ref{fig:avgjsdspermapping} taken only for the single-component systems  with the JSD values averaged over all RDFs per system ($iv$) plots of all  RDFs, potentials, and forces for the SP and EE models for the specific molecules discussed in the main text ($v$) the parameterization method for the new force-fields; and ($vi$) the complete mean-force decomposition plots for the systems shown in Fig.~\ref{fig:meanforcedecomp}. In addition, we provide he list of 3,441 C7O2 isomers used for
the clustering approach in this work as smiles strings, the run files for all of the atomistic
and coarse-grained simulations carried out in this work, including all SP and EE parameters
obtained, and the RDF data for all interactions observed in the 2,476 mapped ensembles
generated in this work. These files can be accessed online.~\cite{zenodo}

\begin{acknowledgments}
  We are grateful to Christoph Scherer for critical reading of the manuscript.
  KHK gratefully acknowledges funding from the Max Planck Graduate
  Center. KHK and TB acknowledge funding from the Emmy Noether program
  of the Deutsche Forschungsgemeinschaft (DFG).
\end{acknowledgments}

\section*{Data Availability}

The data that supports the findings of this study are available within the supplementary material, as well as in a Zenodo repository at http://doi.org/10.5281/zenodo.6032826.~\cite{zenodo}

\bibliography{biblio, references_MPIP, references_PSU}

\end{document}